


\input harvmac.tex
%
%

%
\ifx\epsfbox\UnDeFiNeD\message{(NO epsf.tex, FIGURES WILL BE IGNORED)}
\def\figin#1{\vskip2in}
\else\message{(FIGURES WILL BE INCLUDED)}\def\figin#1{#1}\fi
\def\ifig#1#2#3{\xdef#1{fig.~\the\figno}
\goodbreak\midinsert\figin{\centerline{#3}}%
\smallskip\centerline{\vbox{\baselineskip12pt
\advance\hsize by -1truein\noindent\footnotefont{\bf Fig.~\the\figno:} #2}}
\bigskip\endinsert\global\advance\figno by1}

%
%
\lref\VA{ T. Vachaspati and A. Ach\'ucarro, Phys. Rev. {\bf D44}  (1991) 3067.}
\lref\H{ M. Hindmarsh,  Phys. Rev. Lett. {\bf 68} (1991) 1263.}
\lref\V{ T. Vachaspati, Phys. Rev. Lett.
{\bf 68} (1992) 1977. T. Vachaspati, Nucl. Phys. {\bf B397} (1993) 648.}
\lref\JPV{ M. James, L. Perivolaropoulos and T. Vachaspati,
Nucl. Phys. {\bf B395} (1993) 534.}
\lref\NO{ H. Nielsen and  P. Olesen, Nucl. Phys. {\bf B81} (1973) 45.}
\lref\B{ E. B. Bogomol'nyi, Yad. Fiz. {\bf 24} (1976) 861
[Sov. J. Nucl. Phys. {\bf 24} (1976) 449].}
\lref\AKPV{A. Ach\'ucarro, K. Kuijken, L. Perivolaropoulos and T. Vachaspati,
Nucl. Phys. {\bf B388} (1992) 435.}
\lref\VB{T. Vachaspati and  M. Barriola, ``A New Class of Defects'', preprint
(1992).}
\lref\YN{Y. Nambu,  Nucl. Phys. {\bf B130} (1977) 505.}
\lref\PER{W. B. Perkins, Phys. Rev. {\bf D47}  (1993) 5224.}
\lref\OLE{P. Olesen, ``A $W$-dressed Electroweak String'',
preprint NBI-HE-93-58 (HEP-PH/9310275).}
\lref\AMB{see J. Ambj{\o}rn and P. Olesen, ``Electroweak Magnetism,
$W$-condensation and Anti-Screening'', preprint NBI-HE-93-17 (HEP-PH/9304220),
for a recent review.}

\Title{ \vbox{\baselineskip12pt\hbox{EFI-93-68}\hbox{hep-th/9312034}}}
{\vbox{
\centerline{Cinderella Strings} }}
\centerline{Ana Ach\'{u}carro}
\centerline{\it Department of Mathematics, Tufts University}
\centerline{\it Medford, Massachusetts 02155}
\centerline{\it and}
\centerline{\it Departamento de F\'\i sica Te\'orica, Universidad del Pa\'\i s
Vasco} \centerline{\it Lejona, Vizcaya (Spain)}
\medskip
\centerline{Ruth Gregory}
\centerline{\it Dept. of Applied Mathematics and Theoretical Physics}
\centerline{\it University of Cambridge, Cambridge CB3 9EW (U.K.)}
\medskip
\centerline{Jeffrey A. Harvey}
\centerline{\it Enrico Fermi Institute,University of Chicago}
\centerline{\it 5640 Ellis Avenue,Chicago, Illinois 60637}
\medskip
\centerline{Konrad Kuijken}
\centerline{\it Harvard-Smithsonian Center for Astrophysics}
\centerline{\it 60 Garden Street, Cambridge, Massachusetts 02138}
\bigskip

\noblackbox
\noindent
We investigate recent claims concerning
a new class of cosmic string solutions in the
Weinberg-Salam model. They have the general form of previously
discussed semi-local and electroweak strings, but are modified
by the presence of a non-zero W-condensate in the core of the
string. We explicitly construct such solutions for arbitrary values
of the winding number $N$. We then  prove that they are
gauge equivalent to bare electroweak strings with winding number
$N-1$.
We also  develop new asymptotic expressions for large-$N$ strings.

\Date{November, 1993}

\newsec{Introduction}

There has been much speculation that stable topological defects such
as monopoles, strings, or domain walls may have played an important
role in the early universe. Unfortunately, such defects only arise in
extensions of the standard model which are rather poorly constrained
by experiment.  It  was discovered recently that soliton-like string solutions
can exist in an unphysical version of the Weinberg-Salam model with
the $SU(2)_L$ gauge
coupling set to zero  \VA, \H.
Although strictly speaking these solutions
have no topological charge (the vacuum manifold is simply connected),
they may still be characterized by
a ``winding number'' $N$ which
characterizes the variation of the phase of the Higgs field
as one encircles the string.  This winding is associated with the fact that
the gauge orbit space  is not simply connected.

Semi-local strings also provide
solutions to the physical version of the model with non-zero gauge
coupling \V.  String solutions in the Weinberg-Salam model
had also been discussed earlier in \YN.
For physical values of the coupling constants
the solutions with $N=1$ are unstable \JPV. The linear stability analysis of
\JPV\ does not indicate whether the instability leads to a
dissolution of the
string, or whether it  leads to a modified solution of
lower energy per unit length.  In \PER\ it was shown that the
instability can be viewed as due to  a condensate of  charged
$W^\pm$ vector bosons in the core of the string, a phenomenon
which has been investigated in the context of electroweak symmetry restoration
in magnetic fields \AMB.  In \OLE\  some qualitative
arguments were given suggesting that a modified solution may
exist when $\sin^2 \theta_W = 0$.  As we will show, the perturbation which
seems
most likely to destroy the string has improved stability for winding $N>1$,
suggesting that higher winding strings are a natural place to look for a
modified solution
involving $W$ condensation.
In this paper we will show that  for $N>1$
there is an instability due to $W$ condensation
which leads to an apparently new string solution which
we term the ``Cinderella String'' at  the
physical value $\sin^2 \theta_W \approx 0.23$  when the Higgs mass is less
than the $Z$ mass and
we will construct this  solution numerically. We will then explicitly
construct a gauge transformation that converts this Cinderella string into
an ordinary electroweak string with  winding number one less.

\newsec{Semi-local and Electroweak strings}

The action for the bosonic fields in the Weinberg-Salam model is
\eqn\action{
 S = \int d^4 x
 \left [
         ( D^\mu \Phi )^{\dag}( D_\mu \Phi )  \ -
          {1 \over 4} F_{\mu \nu} F^{\mu \nu} \ -
         {1 \over 4} G_{\mu \nu}^a  G^{a \mu \nu} \
           - \lambda  (\Phi^{\dag} \Phi - \eta^2/2 )^2
     \right ] \ ,
}
where $D_\mu \Phi =
(\nabla_\mu - ig W_\mu ^{\ a}  \sigma^a/2  -   ig'
B_\mu /2 )\Phi $.  $B_{\mu} $  and $W_\mu ^{\ a}$ ($a = 1,2,3$)
are gauge potentials for  $U(1)$ and $SU(2)_L$ respectively,
with field strengths
$F_{\mu \nu}= \partial_{\mu}B_{\nu} - \partial_{\nu}B_{\mu}$
and $ G_{\mu \nu}^a =
\partial_{\mu}W_{\nu} ^{\ a} - \partial_{\nu}W_{\mu}^{\ a}
+ g \epsilon^{abc} W_{\mu}^{\ b} W_{\nu}^{\ c} .$ Our conventions
are such that $\epsilon^{123} = 1$;
$\sigma^a$ are the Pauli matrices. Also,
$ W_\mu^{\ \pm} = (W_\mu^{\ 1} \pm i ~W_\mu^{ \ 2})/\sqrt{2} \ , \
Z_\mu = \cos \theta_w W_\mu^{\ 3} - \sin \theta_w B_\mu \ , \
A_\mu = \sin \theta_w W_\mu^{\ 3} +\cos \theta_w B_\mu $ where
$\sin^2 \theta_w = {g'}^2/\alpha^2 \approx 0.23$ with
$\alpha^2 = g^2 + {g'}^2$. In what follows
we will work in rescaled cylindrical
coordinates $(t,z,r , \theta)$, where $r = \sqrt{\lambda} \eta \rho$
($\rho$ being the usual radial coordinate), and use a coordinate
basis rather than the orthonormal basis often used.

The equations of motion which follow from \action\ are
\eqn\motion{\eqalign{
&D_\mu D^\mu \Phi + 2\lambda \bigl(\Phi^\dagger \Phi - {\eta^2 \over
2} \bigr) \Phi = 0 \cr &\nabla_\mu F^{\mu\nu} + {i g' \over 2}
(\Phi^\dagger D^\nu \Phi - (D^\nu \Phi)^\dagger \Phi ) = 0 \cr &
\nabla_\mu G^{\mu\nu a} + g \epsilon^{abc} W_\mu ^{\ b} G^{\mu\nu c}
+ {i g \over 2} (\Phi^\dagger \sigma^a  D^\nu \Phi - (D^\nu
\Phi)^\dagger  \sigma^a \Phi ) = 0 \cr} }
It was shown in \VA\ for $\theta_w = {\pi  / 2}$ and in \V\
for arbitrary $\theta_w$ that these equations admit a solution
describing a ``Z-vortex":
$ Z_\theta = {2  N( P(r) - 1) / \alpha}, \
\Phi \equiv (\phi_u, \phi_l) = {\eta }( 0,
X(r) e^{ i N\theta} )/\sqrt{2} $ (all other fields set to zero),
 where $X(r)$ and $P(r)$ are the solutions to the
Nielsen-Olesen \NO\ equations
($' \equiv d/dr$):
\eqn\NielsenOlesen{\eqalign {
&- X'' -{ X' \over r} + { N^2  P^2 X\over r^2} + X( X^2 -1) = 0 \cr
& -P'' + { P' \over r} + { 2 \over \beta} P X^2 = 0 \cr}
}
satisfying the boundary conditions $X(0)=0$, $P(0)=1$, and
$ X(r) \to 1, \  P(r) \to 0 $,  as $ r \to \infty $.  The parameter $\beta$ is
the ratio of the square of the scalar and vector masses, $\beta =
{m_{\rm Higgs}^2  / m_Z^2} = { 8 \lambda /  \alpha^2}$.

Note that this is not a ``topological" string in the usual sense.  The vacuum
manifold of the Weinberg-Salam model is $S^3$ and it has no
non-contractible loops. However it has been shown \refs{\VA, \H, \V, \JPV}
that, for certain ranges of values of the Higgs mass and the Weinberg
angle, these configurations are stable to small perturbations.

A special case where stability is easy to understand is $N=1$,
$\cos\theta_w = 0$. In that case the SU(2) gauge
fields decouple ($g=0$) and we have what are called semi-local strings, with
symmetry $SU(2)_{global} \times U(1)_{local}
$ and the same vacuum
manifold  as before. Unlike in the case of  Nielsen-Olesen
strings, where $N=1$ is automatically stable to small perturbations, stability
here is decided by a competition between gradient and potential energy: when
$\beta>1$ there are nearby configurations with a non-zero $\phi_u$ at the core
whose increased gradient energy is more than compensated by a lower
potential energy (from the last term in the action) \H.
For $\beta<1$, however, semi-local strings are
perturbatively stable \refs{\VA, \H, \AKPV \B}. Physically this is because the
variation of the Higgs field at infinity in the $S^3$ vacuum manifold lies on
the orbit of the $U(1)_{local}$ gauge symmetry.  To unwind the string must move
off this orbit, but this raises the energy because the energy due to the
angular dependence of the Higgs field can no longer be canceled
by a $U(1)_{local}$ gauge field.  When  $\cos\theta_w \neq 0$
the barrier to decay becomes finite or may disappear and one must investigate
the
stability numerically.

\newsec{$W$ condensation and large-$N$ vortices}

There are two distinct physical mechanisms that can lower the energy of an
electroweak string. The first one we have just discussed: an instability occurs
if it is  energetically preferable for $\phi_u$ to develop a non-zero value
at the origin. The second one, pointed out in \refs{\V , \PER}, is the
formation of a $W^\pm$ condensate at the core of the string.
 For $N=1$ there is in fact no clean separation of the
known instability into these two types of perturbation. The equations
of motion require a non-zero $\phi_u$ at the origin as soon as there
is a non-zero $W$ condensate. For $N>1$ however there is a clear
distinction between the two.
The first
instability seems likely to
destroy the string, as in the semi-local case, since once the Higgs field moves
away from zero at the origin there is no reason for it not to move back to its
vacuum value everywhere. On the other hand
one might expect the second form of instability
to lead to a new string solution since it does not seem to change the
winding structure in the Higgs field. It was even suggested  in \OLE\
 that a
``$W$-condensed'' solution might be more stable than the  electroweak
string.
In what follows, we investigate these claims.

It can be shown that, inside the core of a large-$N$ vortex,
the Nielsen-Olesen equations \NielsenOlesen\ reduce  at leading
order in $1/N$ to
\eqn\largeno{\eqalign{ -{P}'' + { {P}' \over r} &= 0 \cr
                        { {\xi}' \over \xi} & ={ P \over r}\cr }}
with solution
\eqn\nsol{ P = 1 - p r^2;  \qquad \xi = C r e^{-p r^2/2} }
where $\xi= X^{1/N}$. (The constant $C$ can be estimated by demanding that $\xi
\approx 1$  when $P \approx 0$ which gives $C \approx \sqrt{ p e}$; its precise
value  will not be important in what follows).
In deriving these expressions we have used the fact that $0<\xi<1$ and kept
leading terms in $N$.  We now verify that the $\xi'$ and $\xi''$
terms dropped are indeed subleading by calculating the $N$ dependence
of the coefficient $p$ explicitly.

Since  $X\approx (pe r^2)^{N/2} e^{-Npr^2 / 2}$ in the core, transition from
$X=0$ to $X=1$ occurs fairly abruptly at
$r \approx 1/\sqrt {p}$  with a  width of $\approx 1/\sqrt{Np}$. To study the
transition region we introduce the variables
\eqn\newvar{ u = \sqrt{Np}( r - {1\over \sqrt{p}}); \qquad \gamma =
{1 \over \sqrt{pN}}; \qquad Q = {\sqrt{N} P \over 2 }.}
The leading order behavior of the equations \NielsenOlesen\ then takes the form
\eqn\transit{\eqalign{ -{X}'' + 4X Q^2 + \gamma^2  X
(X^2-1) &= 0  \cr
-{Q}'' + {2 \gamma^2  \over \beta}Q X^2 &=0
\cr}}
with
\eqn\bcs{Q = \cases {-u, & $u \ll 0$ \cr 0& $u \gg 0$
\cr}\quad\quad\quad
X = \cases {0, & $u \ll 0$ \cr 1& $u \gg 0$
\cr}}
The equations \transit\ have a first integral
\eqn\firsti{
-{1 \over 2}({X}')^2 - {\beta \over \gamma^2}({Q}')^2 + 2 X^2
Q^2+ {\gamma^2 \over 4} (X^4 - 2X^2) = {\rm Const}
}
The boundary conditions at  $u = + \infty$ fix the
constant to be $- \gamma^2 /4$,
while the boundary conditions at $u = -\infty$ give
\eqn\findb{
-{ \beta \over \gamma^2}  + {\gamma^2 \over 4} =0
\quad\quad\quad \rightarrow \gamma^2 = 2\sqrt\beta = {1 \over pN}. }
We thus find $p = 1/(2N \sqrt{\beta})$ when $N>>1$. This shows that we were
justified in dropping $\xi'$ and $\xi''$ terms to derive \nsol .
It also implies the large-$N$ scaling of  the  core radius
$r_{\rm core} \approx \sqrt {2N} \beta^{1\over 4}$ (the  width of
the transition region stays of order unity).

Now consider a ($z$- and $\theta$- independent) perturbation
in $\phi_u$
which is
non-vanishing at the origin
\eqn\yinstability{\Phi = {\eta \over \sqrt 2} \pmatrix {f(r) e^{ \Omega t}
 \cr X(r) e^{i N \theta} \cr} , \quad\quad\quad\quad f(r)
\sim f_0 +O(r^2)\quad {\rm as} \ r \to 0}
The corresponding $W$ perturbations decouple if $f(r)$ falls to zero while
inside the core of the string, (since they always appear multiplied by
$X\approx 0$ in the $f$ equation). Inserting the explicit form for $P$ yields
\eqn\feqn{-f'' -{f' \over r} +{ (2\cos^2\theta_w-1)^2\over
4\beta} r^2  f
 + \left({\Omega^2 \over \lambda \eta^2} -1 \right)f = 0  \ .}
The equation can be rescaled to depend
on only one parameter by the change of variables
$ r^2 = {2 \sqrt \beta  \hat{r}^2 /  |2\cos^2\theta_w-1|} $, giving
\eqn\feqnhat{ -f''(\hat{r})  - {f' (\hat{r}) \over \hat{r}} + (\hat{r}^2 -
\Lambda)f(\hat{r}) = 0 \ ,
\quad{\rm where }\quad\Lambda = {2 \sqrt \beta \over
|2\cos^2\theta_w-1|}\left({\Omega^2 \over \lambda \eta^2} - 1\right)}

The solution is $f(\hat{r}) = f_0 e^{-{\hat{r}^2 / 2}}$ and it
corresponds to $\Lambda = -2$ (which is the lowest eigenvalue since f
has no nodes). In principle there could be lower eigenvalues coming from
the other equations (those of the $W$ perturbations) but one can
show that this is not the case. The line of critical stability is
$\Omega = 0$, or
\eqn\betasol{\sqrt\beta =
|2\cos^2\theta_w-1| }
(See Figure 1).
This argument shows that the stability of bare electroweak
strings  with respect
to this particular perturbation is better for large-$N$ strings than for
their $N=1$ counterparts (for $\sin^2 \theta_w = 0.23 $ we find the critical
line at $m_{\rm Higgs} \approx 48.6 GeV$).
Note that consistency requires that $f$ reach zero before
$X$ becomes non-zero; for the physical $\theta_w$ this is true to within
1\%\ when $N> 23$.

\ifig\fone{Stability of electroweak strings to perturbations with
a non-zero $\phi_u$ at the center.
The dashed line shows critical stability for large-$N$ strings to
the specific perturbation of eq.~\yinstability.
The solid line shows critical stability (against {\it all}
perturbations) for the $N=1$ string, in
agreement with  \JPV.}
{\epsfysize=4.5in \epsfbox{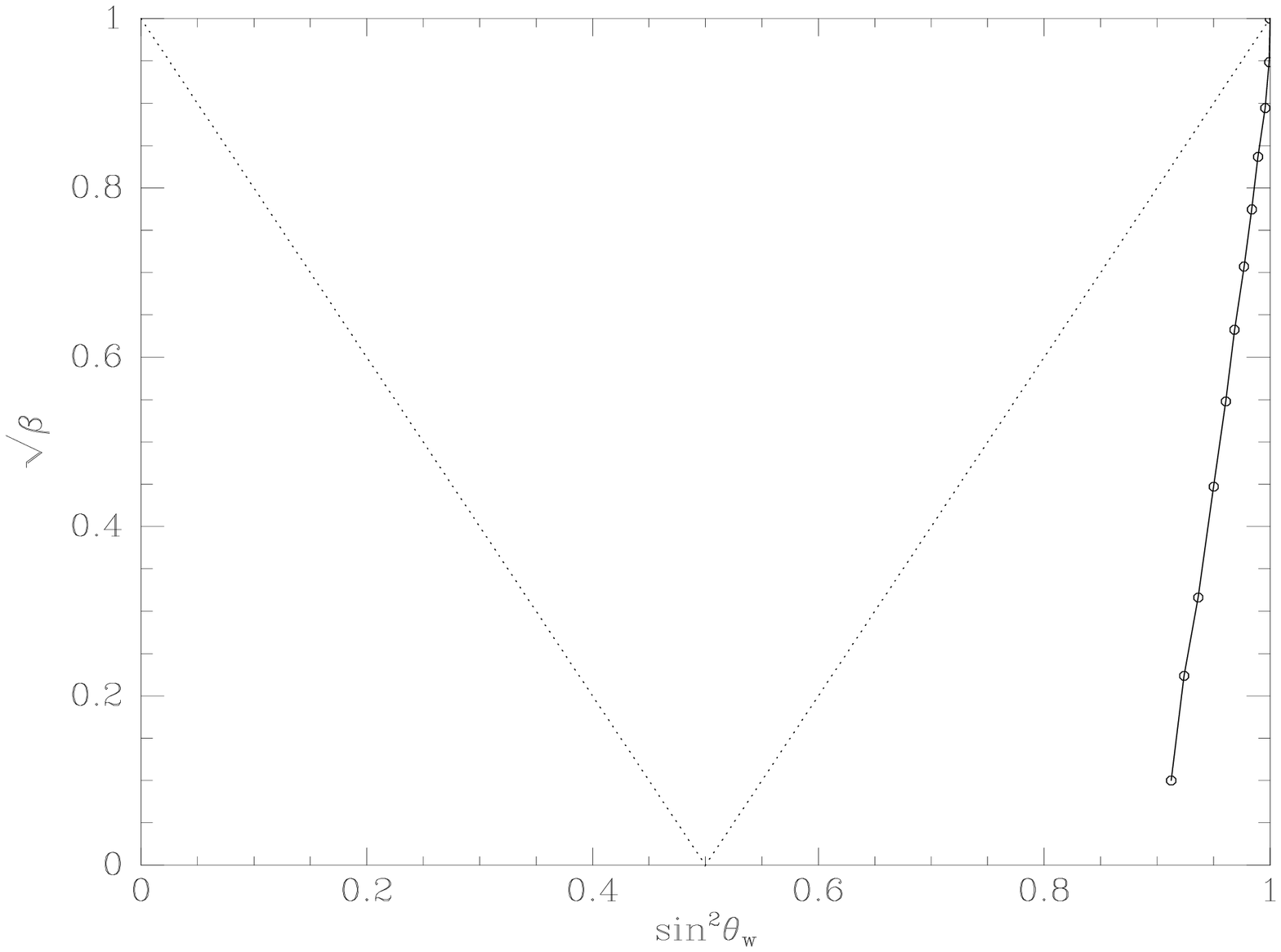}}

We now consider stability with respect to $W$ condensation.
Recall  that inside the string core P only deviates from quadratic  behavior
when X is substantially different from zero.
Thus,  for large $N$ there is a $Z$ magnetic field which is
approximately constant throughout the core with a magnitude given by ${4Np/
\alpha} = 2/ \alpha\sqrt{\beta}$. It is known from  the work in \AMB\ in a
different context that such a uniform magnetic field is unstable to the
formation of a $W$ condensate.  The analysis in \AMB\  shows that such an
instability exists if the magnetic field is greater than a critical value
determined by the value of the $W$ mass. Since the $W$ fields are essentially
massless in the core of the vortex at large $N$, the field in the core is large
enough to cause this instability.  The static energy is minimized by having
a condensate with a  constant $W$ magnetic moment in the core.
In cylindrical coordinates this suggests an ansatz of the form
\eqn\wansatz{
W_r^\pm = \pm i{\sqrt\lambda}\eta v(r) e^{i \nu \theta}; \qquad
\ W_\theta^\pm = m(r) e^{i \nu \theta} ,}
where $\nu $ is a {\it positive}
integer. An extension of
the analysis in \PER\ for arbitrary $N$ shows that an efficient way to
lower the energy which is compatible with the equations of motion is to have
$\nu = 1$
and the leading $r$ behavior given by
$v(r) = v_0 + O(r^2)$, $m(r) =- v_0 r+ O(r^3)$.
\foot{ Note that the angular dependence of the
condensate differs by a sign from \OLE\ and
the radial dependence is one power of $r$ less than that
given in  \PER\ for $N=1$.}

Moreover, analysis of the equations of motion with this angular
dependence shows that it is not
consistent to set the upper component of the Higgs field, $\phi_u$,  to zero.
An upper component
of the form $Y(r)e^{i(N-1) \theta}$ is required, with $Y(r) =y_0r^{N-1} +
O(r^{N+1})$. \foot{ For general $\nu \leq N$ we
find $Y \sim e^{i(N-\nu)\theta}
r^{N - \nu}$ as $r \to 0$.}

We thus see that at large $N$ the string has improved stability with
respect to perturbations of $\phi_u$ and has an instability due
to $W$ condensation. In the following section we follow numerically
the instability due to $W$ condensation to find a modified string solution.

\newsec{A $W$ dressed solution}

Following the previous discussion we consider the following
ansatz for a $W$ condensed solution
\eqn\ansatz{\eqalign{\Phi \equiv \left( \phi_u,\phi_l \right) & =
{\eta \over \sqrt{2}} \left( Y(r) e^{i(N-1) \theta},
                      X(r) e^{i N \theta} \right)  \cr
                      W_r &= i \sqrt{\lambda} \eta v(r) e^{i \theta};
                \qquad\quad W_\theta = m(r) e^{i \theta} \cr
                Z_\theta &= {2 N \over \alpha} (P(r)-1); \qquad
                A_\theta = {2 N \over \alpha}A(r) . \cr }}
With this ansatz the energy is given by
\eqn\energize{\eqalign{
E &= {\eta^2 \over 2 }\int r\ dr \Biggl\{  (Y' - {gvX \over \sqrt 2})^2 +
(X' + {gvY \over \sqrt 2})^2  + {1 \over 2} (X^2 + Y^2 - 1)^2\cr
&  + { 1 \over r^2} (NPX - {gmY \over \sqrt 2})^2
+ {\beta\alpha^2\over 4r^2}\left[ m' + v\left( 1 + {2g^2\over \alpha^2}N(P-1)
+ {2gg'\over \alpha^2}NA \right) \right]^2 \cr
& - {\beta \over r^2} [ g^2 NP' + gg'NA']vm
+ {\beta \over 2} \left[  \left({NP' \over r}\right)^2
+ \left({NA' \over r}\right)^2 + { g^2 \alpha^2 v^2 m^2 \over r^2}\right]
\cr
& + {1 \over r^2} \left[ (N-1) Y -
\left({g^2-g'^2\over\alpha^2}\right) N(P-1)Y -
{2g g'\over \alpha^2}NAY - {gmX
\over \sqrt 2}\right]^2  \ \Biggr\}. \cr }}

To find a solution of the equations of motion \motion\ we used a variational
method to look for minima for the energy \energize\ given the ansatz
\ansatz. Each of the fields $X^{1/N}$, $Y^{1/(N-1)}$, $P$, $A$, $m$ and $v$ in
the energy was written as a finite Fourier series running from $r=0$ to a
sufficiently large outer radius $r_{\rm out}$ (typically twice the core
radius), and the harmonics' coefficients varied until the energy was minimized.
The harmonic series were constrained to satisfy the boundary conditions
$X^{1/N}\sim r$, $Y^{1/(N-1)}\sim r$, $A\sim r^2$, $m\sim r$, $v\sim {\rm
const}$, $P-1\sim r^2$ at $r=0$,
and $P$ was forced to be zero at radius $r_{\rm
out}$. The minimum energy configuration was then refined by increasing the
number of harmonics, until the energy stayed constant to one part in $10^5$.
 We then substituted these configurations into the equations
of motion \motion\ to confirm that we had found a solution.  Working with $15$
basis functions for each function and using $60$ discrete radial points in the
integrations, we found that the equations of motion were satisfied except
for high-frequency residuals. In figure~2 we compare, as a function of $\beta$,
the energies per unit length of
the $W$-condensed solutions with $N=2,4$ and of bare
electroweak strings with $N=1,2,3,4$.

\ifig\ftwo{The energy per unit length for the electroweak strings
(solid lines) as a function of $\beta$ and $N$, and for the Cinderella
strings with $N=2$ and 4 (dots).}{\epsfysize=4.5in\epsfbox{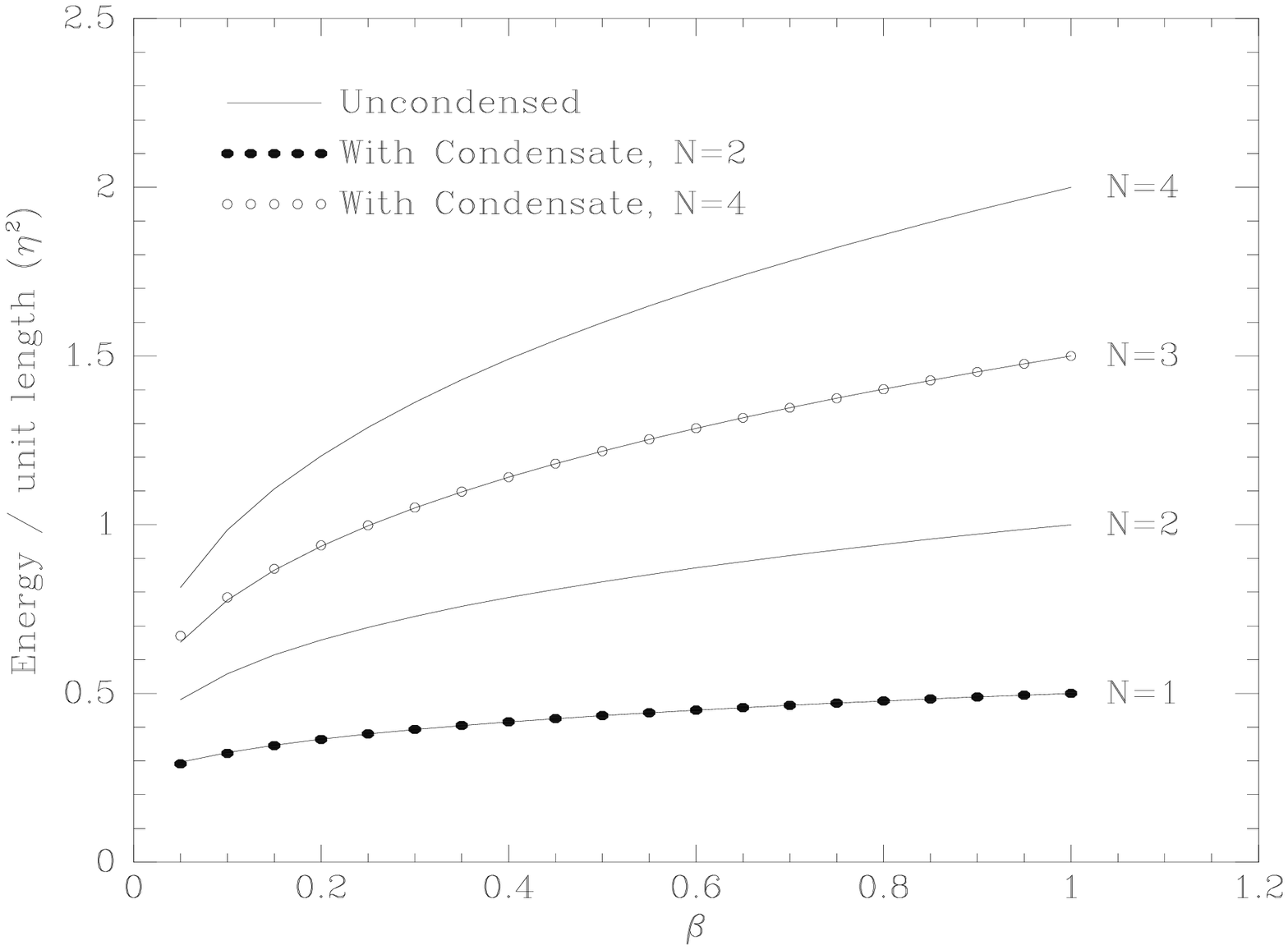}}

We have also looked for solutions that minimize the energy for $N=1$.
For the range of $\beta$ and $\sin^2 \theta_w$ shown in figure~1 we found
that the bare electroweak string was a solution and was stable to the
development of a $W$ condensate. Our results are in agreement with
those of \JPV.  For values of $\beta$ and
$\sin^2 \theta_w$ outside this range the program would only converge
to a vacuum solution, consistent with the lack of existence of a
non-trivial $W$ condensed electroweak string with $N=1$.

\newsec{The clock strikes twelve}
Inspection of the energy per unit length of the Cinderella string
with winding $N$ and the bare electroweak string with winding $N-1$
shows that they are the same to our numerical accuracy. This strongly
suggests that they are in fact the same solution up to a
gauge transformation. To show that this
is the case we construct a gauge transformation relating the two solutions
and verify numerically that they are in
fact equivalent. To construct the gauge transformation it is simplest
to first focus on the Higgs field. For the Cinderella string it takes
the form given by \ansatz\ with asymptotic behavior
\eqn\asymp{\eqalign{X(r) & \rightarrow 1, \qquad  \quad Y(r) \rightarrow 0,
				\qquad \qquad\ \ r \rightarrow \infty \cr
           X(r) & \rightarrow x_0 r^N , \quad Y(r) \rightarrow y_0 r^{N-1} ,
           \qquad r \rightarrow 0. \cr }}
This may be put in the form of the Higgs field for a bare electroweak
string of winding $N-1$, ${\tilde \Phi} =  {\bf G}\, \Phi$  by the gauge
transformation
\eqn\gtransf{ {\bf G} = \pmatrix{ \cos \psi e^{i \theta} & -\sin \psi \cr
                                 \sin \psi & \cos \psi e^{-i \theta} \cr }}
where
\eqn\psidef{ \cos \psi(r) = {X(r) \over \sqrt{X^2(r) + Y^2(r)}}}
We have verified numerically that the resulting Higgs field profile is
the same as that for a bare electroweak string of winding $N-1$.
We have also checked that
the gauge transformed gauge fields,
\eqn\newws{\eqalign{ {\tilde W}_r &= i \sqrt{2}
                    \left( v(r) -{ \sqrt{2} \over g} \psi'  \right)
                    e^{-i \theta} \cr
                    {\tilde W}_\theta &= \sqrt{2} \left(
                    m(r) \cos 2 \psi + (W^3_\theta(r) +{1 \over g} )
                    {\sin 2 \psi \over \sqrt{2} }\right) e^{-i \theta} \cr
                    {\tilde W}^3_\theta &= W^3_\theta(r) \cos 2 \psi -
                    \sqrt{2} m(r) \sin 2 \psi + {2 \over g} \cos^2 \psi , \cr
}}
are those of an electroweak string with winding $N-1$, that is,
${\tilde W}_r={\tilde W}_\theta={\tilde A}_\theta=0$ and
\hbox{${\tilde Z}_\theta = 2(N-1)(P(r)-1)/\alpha$} with $P(r)$ the
Nielsen-Olesen solution for winding $N-1$.

\newsec{Conclusions}
We have shown that for the physical value
of  $\sin^2 \theta_w$ electroweak strings with winding $N>1$ have
an instability associated with the formation of a condensate of
$W^\pm$  vector bosons in the core of the string. By minimizing the energy with
an ansatz compatible with this condensate we were able to find a $W$ condensed
Cinderella string.  However this new Cinderella string is an illusion, it is
gauge equivalent to a bare electroweak string with winding $N-1$.

By following the $W$ condensate instability numerically to a new
solution we have shown that there are no new $W$ condensed electroweak
string solutions within the ansatz we have used.  While we cannot
completely rule out the possibility of more complicated stable string
solutions in the Weinberg-Salam model, our results make it quite
likely that none exist. This leaves only the $N=1$ electroweak string
which, as has been shown in a previous analysis \JPV\ and reconfirmed
here, is not stable for physical values of the Weinberg angle.

\centerline{\bf Acknowledgements}
This work was supported in part by NSF grants PHY90-00386 (J.H.) and
PHY91-11188 (A.A.), by the McCormick Fund of the Enrico Fermi
Institute  and the SERC (R.G.), by Hubble Fellowship grant HF-1020.01-91A
(K.K.),
and by CICYT grant AEN-90-0330 (A.A.). J.H.  also acknowledges support
from NSF PYI grant PHY-9157463.  A.A.~thanks the Enrico Fermi
Institute, and K.K.~the University of the Basque Country, for their
hospitality.

\listrefs

\bye